\documentclass[a4paper]{PoS}

\title{Rare \emph{B} decays with moving NRQCD and improved staggered quarks}

\ShortTitle{Rare B decays with moving NRQCD and improved staggered quarks}

\author{\speaker{Stefan Meinel}$\:^a$, Eike H. M\"uller$^b$, Lew Khomskii$^a$,
        Alistair Hart$^b$, Ronald R. Horgan$^a$, Matthew Wingate$^a$\\ \\
\llap{$^a$}DAMTP, University of Cambridge, Wilberforce Road, Cambridge CB3 0WA, UK\\
\llap{$^b$}SUPA, School of Physics and Astronomy, University of Edinburgh,
           Edinburgh EH9 3JZ, UK\\ \\
E-mail: \email{S.Meinel@damtp.cam.ac.uk}}

\abstract{We calculate form factors relevant for rare $B$ decays using moving-NRQCD
for the $b$ quark and the AsqTad action for the light quarks. Moving NRQCD allows us
to work directly with the physical $b$ quark mass and go to higher recoil momentum
compared to standard NRQCD. Here, we show first results for the matrix elements and
the operator matching coefficients. Some difficulties and possible ways of improvement
are discussed.
}

\FullConference{The XXVI International Symposium on Lattice Field Theory \\
		 July 14 - 19, 2008\\
		 Williamsburg, Virginia, USA}

\begin{document}

\section{Introduction}

Decays of $B$ mesons via the flavour-changing-neutral-current transition $b \rightarrow s$
are particularly sensitive to possible new-physics contributions and provide tests of
the CKM mechanism at the loop level. Measurements of exclusive modes like
$B\rightarrow K^* \gamma$ have reached a good accuracy, and call for precise theoretical
predictions. These are more difficult than for tree-level decays such as
$B\rightarrow \pi l \nu$, since a large set of effective electroweak operators contributes
and long-distance or spectator effects can be important. Nevertheless, the computation
of hadronic form factors in lattice QCD is highly desirable and complements continuum
approaches.

\begin{table}[ht!]
\begin{center}
\begin{tabular}{ccl}
\hline
Matrix element & Form factor & Relevant decay(s) \\ \hline 
$\langle P|\bar{q}\gamma^\mu b|B\rangle$ & $f_+, f_0$ & 
$\left\{\begin{array}{l} B\to\pi\ell\nu\\B\to K\ell^+\ell^-\end{array}\right.$
\\[4mm]
$\langle P|\bar{q}\sigma^{\mu\nu}q_\nu b|B\rangle$ & $f_T$ &
\hspace{3.5mm}$B\to K\ell^+\ell^-$ \\[2mm]
$\begin{array}{c}\langle V|\bar{q}\gamma^\mu b|B\rangle
\\ \langle V|\bar{q}\gamma^\mu\gamma^5 b|B\rangle\end{array}$ & 
$\begin{array}{c}V\\ A_0, A_1, A_2\end{array}$ &
$\left\{\begin{array}{l} B\to(\rho/\omega)\ell\nu \\
B\to K^*\ell^+\ell^-\end{array}\right.$ \\[5mm]
$\begin{array}{c}\langle V|\bar{q}\sigma^{\mu\nu}q_\nu b|B\rangle \\
\langle V|\bar{q}\sigma^{\mu\nu}\gamma^5 q_\nu b|B\rangle\end{array}$ &
$\begin{array}{c}T_1\\T_2, T_3\end{array}$ &
$\left\{\begin{array}{l} B\to K^*\gamma \\
B\to K^*\ell^+\ell^-\end{array}\right.$ \\ \hline
\end{tabular}
\end{center}
\vspace{-2ex}
\caption{Form factors for semileptonic and radiative $B$ decays.}
\label{tab:formfact}
\end{table}

We are currently working on the calculation of the form factors listed in
Table~\ref{tab:formfact}. The combination of NRQCD and improved staggered
actions for heavy-light mesons has already proven very successful in the
calculation of form factors \cite{Dalgic:2006dt}. In order to extend the
kinematic range to high recoil (lower $q^2$), we now use a moving-NRQCD
(mNRQCD) action for the heavy quark. A brief discussion of our strategy
can be found in \cite{Meinel:2007eh}, and a new detailed account of mNRQCD
will be given in \cite{mNRQCDpaper}. Here, we report on the progress in the
computation of matrix elements and operator matching coefficients achieved
so far.

\section{Lattice methods}

The matrix element $\langle F(p')| J | B(p)\rangle$, where $F$ denotes
the final pseudoscalar ($P$) or vector ($V$) meson and $J$ is the relevant
current in the effective electroweak operator (see Table~\ref{tab:formfact}),
can be extracted from the combination of the Euclidean 3-point correlator
\begin{equation}
C_{FJB}(\mathbf{k}_{(\mathbf{q})},\:\mathbf{k}_{(\mathbf{p})},\:x_0,\:y_0,\:z_0)
=\sum_{\mathbf{y}}\sum_{\mathbf{z}}\left\langle \Phi_F(x)\:J^{(\mathrm{lat})}(y)
\:\Phi_B^\dag(z)\right\rangle e^{-i\mathbf{p'}\cdot\mathbf{x}}
e^{-i\mathbf{k}_{(\mathbf{q})}\cdot\mathbf{y}}
e^{i\mathbf{k}_{(\mathbf{p})}\cdot\mathbf{z}} \label{eqn:b3pt}
\end{equation}

\null\vspace{-6.7ex}\null

\noindent with the two-point functions

\null\vspace{-9.7ex}\null

\begin{eqnarray}
C_{BB}(\mathbf{k}_{(\mathbf{p})},\:x_0,\:y_0)&=&\sum_{\mathbf{x}}
\left\langle \Phi_B(x)\:\Phi_B^\dag(y)\right\rangle
e^{-i\mathbf{k}_{(\mathbf{p})}\cdot(\mathbf{x}-\mathbf{y})},\\
C_{FF}(\mathbf{p'},\:x_0,\:y_0)&=&\sum_{\mathbf{x}}
\left\langle \Phi_F(x)\:\Phi_F^\dag(y)\right\rangle
e^{-i\mathbf{p'}\cdot(\mathbf{x}-\mathbf{y})}.
\end{eqnarray}

\null\vspace{-6ex}\null

\noindent Here, $\Phi_B$ and $\Phi_F$ are suitable interpolating fields for the
initial and final meson, and $J^{(\mathrm{lat})}$ is a lattice version of the
current, obtained by operator matching (see section \ref{sec:matching}). For the
light quarks, we convert from 1-component staggered to 4-component \emph{naive}
fields \cite{Wingate:2002fh}. Due to the use of mNRQCD for the $b$ quark,
the physical momenta $\mathbf{p}$ and $\mathbf{q}$ are related to the lattice
momenta $\mathbf{k}_{(\mathbf{p})}$ and $\mathbf{k}_{(\mathbf{q})}$ by \vspace{-0.5ex} \null
\begin{eqnarray}
\nonumber\mathbf{p}&=&\mathbf{k}_{(\mathbf{p})}+Z_p\:\gamma\: m_b \mathbf{v},\\
\mathbf{q}&=&\mathbf{k}_{(\mathbf{q})}+Z_p\:\gamma\: m_b \mathbf{v}
\end{eqnarray}
where $Z_p\approx 1$ is the renormalisation of the external momentum,
$\gamma =1/ \sqrt{1-\mathbf{v}^2}$ and $\mathbf{v}$ is the boost velocity. One has
$\mathbf{p'}=\mathbf{p}-\mathbf{q}=\mathbf{k}_{(\mathbf{p})}-\mathbf{k}_{(\mathbf{q})}$.
The physical energy $p_0=E_B$ of the $B$ meson is also shifted, \vspace{-3ex} \null
\begin{eqnarray}
E_B(\mathbf{p})=E_\mathbf{v}(\mathbf{k}_{(\mathbf{p})})+\Delta_\mathbf{v} \label{eq:eshift}
\end{eqnarray}
where $E_\mathbf{v}(\mathbf{k}_{(\mathbf{p})})$ is the unphysical energy obtained
from the fit to the correlator and $\Delta_\mathbf{v}$ is the velocity-dependent
energy shift. Writing $t=|x_0-y_0|$ and $T=|x_0-z_0|$, the correlators are fitted by
\begin{eqnarray}
C_{FJB}(\mathbf{k}_{(\mathbf{q})},\:\mathbf{k}_{(\mathbf{p})},\:t,\:T)&
\rightarrow&\sum_{k=0}^{K-1}\:\:\sum_{l=0}^{L-1}A_{kl}^{(FJB)}
(-1)^{k\:t}(-1)^{l(T-t)}e^{-E'_k t} e^{-E_l(T-t)},\label{eq:3ptfit}\\
C_{BB}(\mathbf{k}_{(\mathbf{p})},\:t)&\rightarrow&
\sum_{l=0}^{L-1}A_l^{(BB)}(-1)^{l(t+1)}e^{-E_l t},\\
C_{FF}(\mathbf{p'},\:t)&\rightarrow&16
\sum_{k=0}^{K-1}A_k^{(FF)}(-1)^{k(t+1)}e^{-E'_k t}
\end{eqnarray}
or equivalent parametrisations. Every other exponential comes with an oscillating
pre-factor, as required by the use of naive quarks \cite{Wingate:2002fh}. The
correlator $C_{FF}$ receives an extra factor of 16 due to the trace over a $16\times16$
taste matrix, while the heavy-light correlators $C_{BB}$ and $C_{FJB}$ receive
contributions from only one taste \cite{Wingate:2002fh}. The ground-state fit
parameters are related to the matrix elements as follows:
\begin{eqnarray}
A_{00}^{(FJB)}&=&\left\{\begin{array}{ll}\displaystyle\frac{\sqrt{Z_V}}{2E_V}
\frac{\sqrt{Z_B}}{2E_B}\sum_s\varepsilon_j(p',s)
\:\langle V\left(p',\varepsilon(p',s)\right)| \:J \:| B(p)\rangle, & F=V,\\
\displaystyle\frac{\sqrt{Z_P}}{2E_P}\frac{\sqrt{Z_B}}{2E_B}
\:\langle P\left(p'\right)| \:J \:| B(p)\rangle, & F=P \end{array} \right.\label{eq:3pt_ampl}\\
A_0^{(BB)}&=&\frac{Z_B}{2E_B} \label{eq:Z1}\\
A_0^{(FF)}&=&\left\{\begin{array}{ll}\displaystyle\sum_s\frac{Z_V}{2E_V}
\:\varepsilon^*_j(p',s)\varepsilon_j(p',s), & F=V, \\
\displaystyle\frac{Z_P}{2E_P} , & F=P. \end{array} \right. \label{eq:Z2}
\end{eqnarray}
The amplitudes $\sqrt{Z_B}$, $\sqrt{Z_P}$ and $\sqrt{Z_V}$ in (\ref{eq:3pt_ampl})
depend on the form of the interpolating fields $\Phi_B$, $\Phi_P$ and $\Phi_V$ and
can be extracted from (\ref{eq:Z1}) and (\ref{eq:Z2}).

\section{Operator matching}

\label{sec:matching}

The continuum currents $J$ must be replaced by lattice currents $J^{(\mathrm{lat})}$
containing suitable matching coefficients to correct for the different ultraviolet
behaviour of QCD and lattice mNRQCD. As only the high-energetic modes with
$E \gtrsim m_b$ differ in the theories and $\alpha_s(m_b) \ll 1$, matching coefficients
can be computed perturbatively.

We use tadpole-improved 1-loop lattice perturbation theory. The Feynman rules are
generated automatically \cite{Hart:2004bd} and diagrams are evaluated using the
Monte Carlo integrator VEGAS.

The first step is the computation of a set of heavy-quark renormalisation parameters
from the self-energy diagrams: the zero-point energy $E_0$, the wavefunction
renormalisation $Z_\psi$, the renormalisation of the mass $Z_m$ and the renormalisation
of the boost velocity $Z_v$ \cite{Dougall:2004hw, mNRQCDpaper}. Results for the full
improved $\mathcal{O}(\Lambda_{QCD}^2/m_b^2)$ lattice mNRQCD action will be presented
in \cite{mNRQCDpaper}.

Once these parameters are known, one can proceed with the calculation of matching
coefficients. For the (axial-)vector currents, these have been computed in the static
limit (i.e. neglecting $\mathcal{O}(\Lambda_{QCD}/m_b)$ corrections in
$J^{(\mathrm{lat})}$), and the calculation including the $\mathcal{O}(\Lambda_{QCD}/m_b)$
corrections is underway \cite{LewThesis}. In the following, we focus on the tensor
current, which, in the continuum, is given by\footnote{For the matching calculation,
we treat the light quark as massless; in this limit the matching coefficients for the
last two operators in Tab.~\ref{tab:formfact} are equal due to chiral symmetry.}
\vspace{-2.5ex} \null
\begin{eqnarray}
J_7^{\mu\nu} &=& \frac{e}{16\pi^2} m_b \;\overline{q} \sigma^{\mu\nu} b
\qquad\mathrm{with}\quad\sigma^{\mu\nu} = \frac{i}{2}[\gamma^\mu,\gamma^\nu].
\end{eqnarray}
We work in the static limit. At this order in the heavy-quark expansion there are
two operators with different Dirac structure in lattice mNRQCD. For the $\mu=0$
components one has \vspace{-0.5ex} \null
\begin{eqnarray}
\nonumber J_{7,1}^{0\ell} &=& -\frac{e}{16\pi^2}m_b\sqrt{\frac{\scriptstyle 1+\gamma}{\scriptstyle 2\gamma}}
\left(\overline{q} \sigma_{0\ell}\Psi_v^{(+)}\right),\\ 
J_{7,2}^{0\ell} &=& i\frac{e}{16\pi^2}m_b\: v\sqrt{\frac{\scriptstyle \gamma}{\scriptstyle 2(1+\gamma)}}
\left(\overline{q} \sigma_{0\ell}\vec{\hat{v}}\cdot\vec{\gamma}\gamma_0\Psi_v^{(+)}\right)
\end{eqnarray}

\begin{figure}[ht!]
\begin{minipage}[t]{.41\linewidth}
\centerline{\includegraphics[scale=0.8]{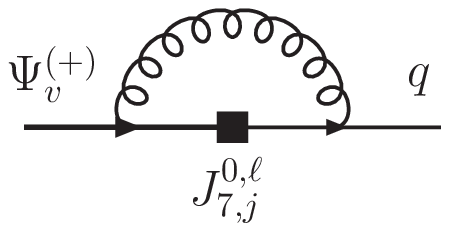}}
\caption{Vertex correction diagram.}
\label{fig:tensor_matching_diagram}
\end{minipage}
\hfill
\begin{minipage}[t]{.57\linewidth}
\centerline{\includegraphics[scale=0.28]{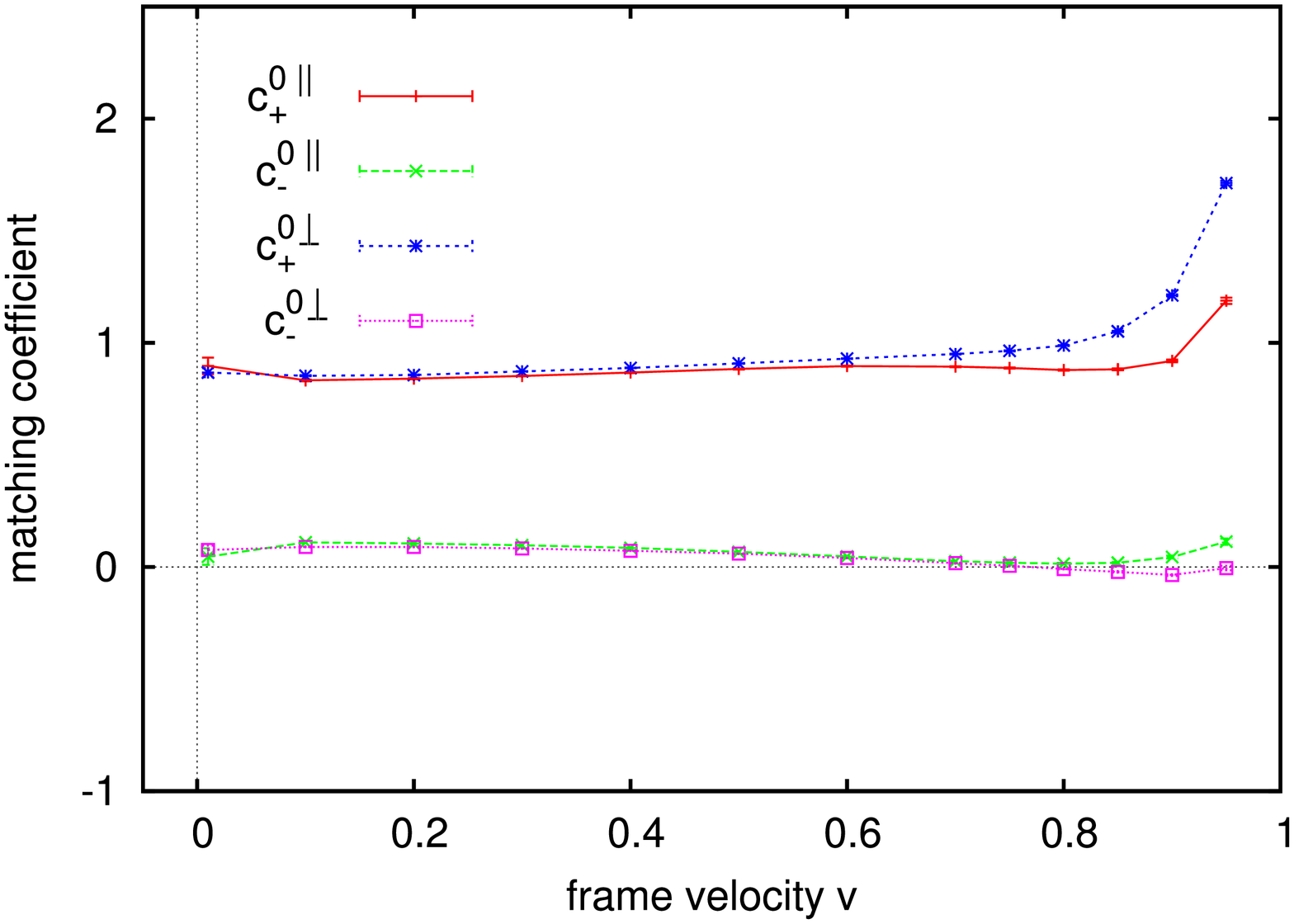}}
\vspace{-1.5ex}
\caption{Matching coefficients for the tensor current. The subtraction point is $\mu=1/a$.}
\label{fig:tensor_matching_coefficients}
\end{minipage}
\end{figure}

\null\vspace{-5.6ex}\null

\noindent where $\Psi_v^{(+)}$ denotes the mNRQCD field with the antiquark components set to zero.
On the lattice, these operators mix under renormalisation; the one-particle irreducible vertex
correction that contributes in the static limit is shown in Fig. \ref{fig:tensor_matching_diagram}.

\vspace{-3ex}

\noindent Writing $J_{7,\pm}^{0\ell} = J_{7,1}^{0\ell} \pm J_{7,2}^{0\ell}$, we obtain the
lattice operator
\begin{eqnarray}
\nonumber J_7^{(\mathrm{lat})0\ell} &=& (1+\alpha_s c_1^{0\ell}) J_{7,1}^{0\ell}
+ (1+\alpha_s c_2^{0\ell}) J_{7,2}^{0\ell}\\
&=& (1+\alpha_s c_+^{0\ell}) J_{7,+}^{0\ell} + \alpha_s c_-^{0\ell} J_{7,-}^{0\ell}.
\end{eqnarray}
The matching coefficients $c_{\pm}^{0\ell}$ must be adjusted such that
$J_7^{(\mathrm{lat})0\ell}$ has the same one-loop matrix elements as the tensor operator
in the continuum theory. They depend on the lattice spacing and contain a logarithmic
ultraviolet divergence as the tensor operator is not conserved.

Results using an improved $\mathcal{O}(\Lambda_{QCD}/m_b)$ mNRQCD action at $am=2.8$,
$n=2$, the AsqTad action for the light quark and the L\"{u}scher-Weisz gluon action
are shown in Fig. \ref{fig:tensor_matching_coefficients}. The matching coefficient of
the operator $J_{7,-}^{0\ell}$, which only arises at 1-loop level, is strongly suppressed
and the dependence on the frame velocity is found to be small for all matching coefficients.

\vspace{-1ex}

\section{Details of the numerical calculations and first results}

\vspace{-1ex}

In our first computations of the 3-point functions (\ref{eqn:b3pt}) we used the local
interpolating fields $\Phi_B(z)=\bar{q}'(z)\gamma_5b(z)$ and
$\Phi_F(x)=\bar{q}'(x)\Gamma_Fq(x)$ with $\Gamma_P=\gamma_5,\:\Gamma_V=\gamma_{1,2,3}$.
As in standard NRQCD, the heavy-quark Green function $G_b(y,z)$ can be obtained by
solving an initial value problem. Let us consider the case $x_0>y_0>z_0$. Schematically,
as initial value at $z_0$ we use the propagator of the light valence quark,
$\gamma_5\:G_{q'}(z, x)$, and then evolve the heavy-quark Green function up to the
time slice $y_0$ where we perform the contraction with the various gamma matrices and
the other light-quark propagator $G_q(x, y)=\gamma_5G_q^\dag(y, x)\gamma_5$. This method
only requires light-quark propagators with a fixed origin $x$, and, since the current is
inserted only in the final contraction, allows the efficient simultaneous computation of
arbitrary currents.

These initial calculations were done on 400 MILC gauge configurations of size
$20^3\times64$ with 2+1 flavours of light quarks, at $\beta=6.76$ and $a^{-1}\approx1.6$
GeV. The light sea quark masses were $am_u=am_d=0.007$, $am_s=0.05$ and the light valence
quark masses $am_u=am_d=0.007$, $am_s=0.04$ (we used the AsqTad action). On each
configuration, we took four different origins $x$, and additionally averaged with the
time-reversed process.

Even though we have implemented the full $\mathcal{O}(\Lambda_{QCD}^2/m_b^2)$ mNRQCD
action, we only used an $\mathcal{O}(\Lambda_{QCD}/m_b)$ mNRQCD action here to
save computer time. This is sufficiently accurate since we only considered currents in the
static limit here. The heavy-quark mass was set to $am_b=2.8$ and the stability parameter
was $n=2$. All lattice momenta and the boost velocity were always pointing in 1-direction.
In this case, 21 combinations of operators/indices and final-state polarisations give
non-zero contributions, and all the form factors listed in Table \ref{tab:formfact}
can be extracted from them.

We performed Bayesian multi-exponential fits in the two variables $T$ and $t$. Gaussian
priors for the ground state energies were taken from fit results of the corresponding
two-point functions, with widths equal to the error from the fit result. The mNRQCD
energy shift $\Delta_\mathbf{v}$ (see eq. (\ref{eq:eshift})) was determined
non-perturbatively from heavy-heavy meson dispersion relations (for those, the full
mNRQCD action accurate to $v_{nr}^4$ in heavy-heavy power counting was used). A bootstrap analysis
was used to determine the form factors and their statistical errors.

\begin{figure}[ht!]
\begin{minipage}[t]{.48\linewidth}
\centerline{\includegraphics[width=\linewidth]{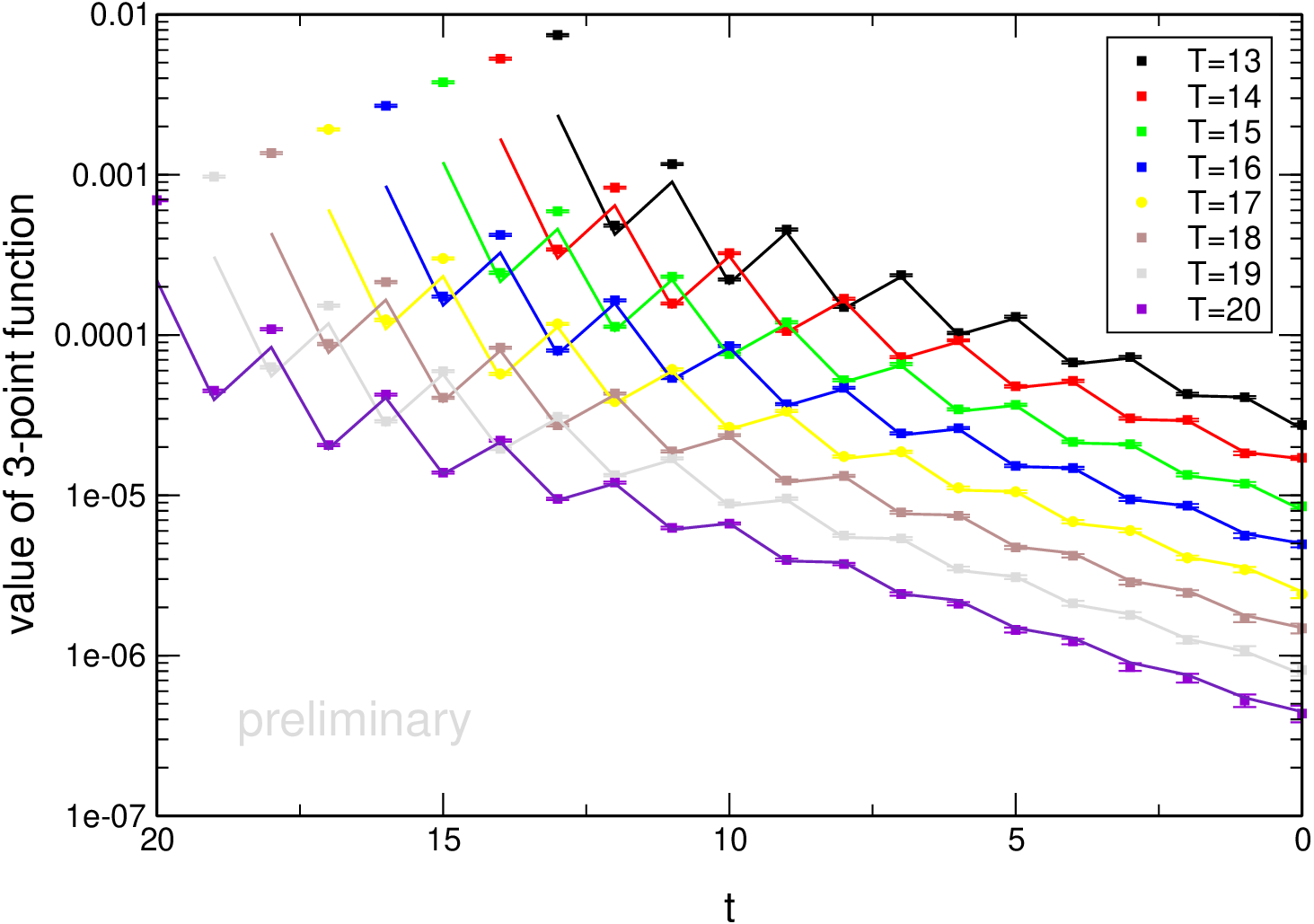}}
\caption{Three-point correlator $\langle \:\:\Phi_K \:\:\: \bar s \gamma_0 b \:\:\:
\Phi_B^\dag\:\: \rangle$ \footnotesize at $\mathbf{k}_{(\mathbf{p})}=0$,
$\mathbf{k}_{(\mathbf{q})}=0$, $\mathbf{v}=0$. The fitting range is $T=14\:...\:18$
and $t=6\:...\:(T-5)$.}
\label{fig:g5_g0_sl_q_0_0_0_p_0_0_0_v_00}
\end{minipage}
\hfill
\begin{minipage}[t]{.48\linewidth}
\centerline{\includegraphics[width=\linewidth]{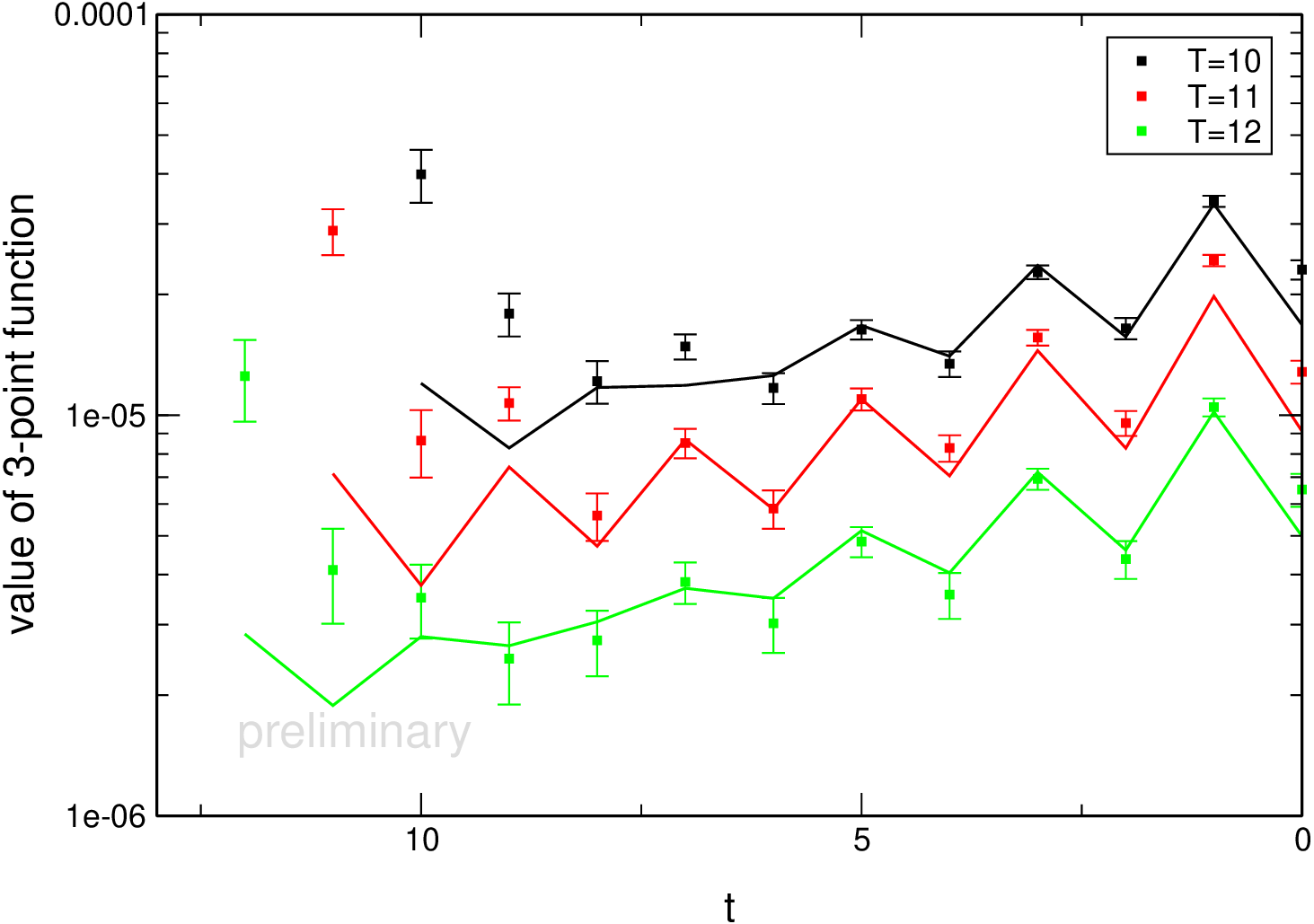}}
\caption{Three-point correlator  $\langle \:\:\Phi_{K^*} \:\:\: \bar s \sigma_{1 3} b
\:\:\: \Phi_B^\dag\:\: \rangle$\footnotesize at $\mathbf{k}_{(\mathbf{p})}=0$,
$\mathbf{k}_{(\mathbf{q})}=\frac{2\pi}{L}(1,0,0)$, $\mathbf{v}=0$. The fitting range
is $T=8\:...\:20$ and $t=4\:...\:(T-4)$ (not all data shown for legibility).}
\label{fig:g2_s13_sl_q_1_0_0_p_0_0_0_v_00}
\end{minipage}
 \end{figure}

\begin{figure}[ht!]
\begin{minipage}[t]{.48\linewidth}
\centerline{\includegraphics[width=\linewidth]{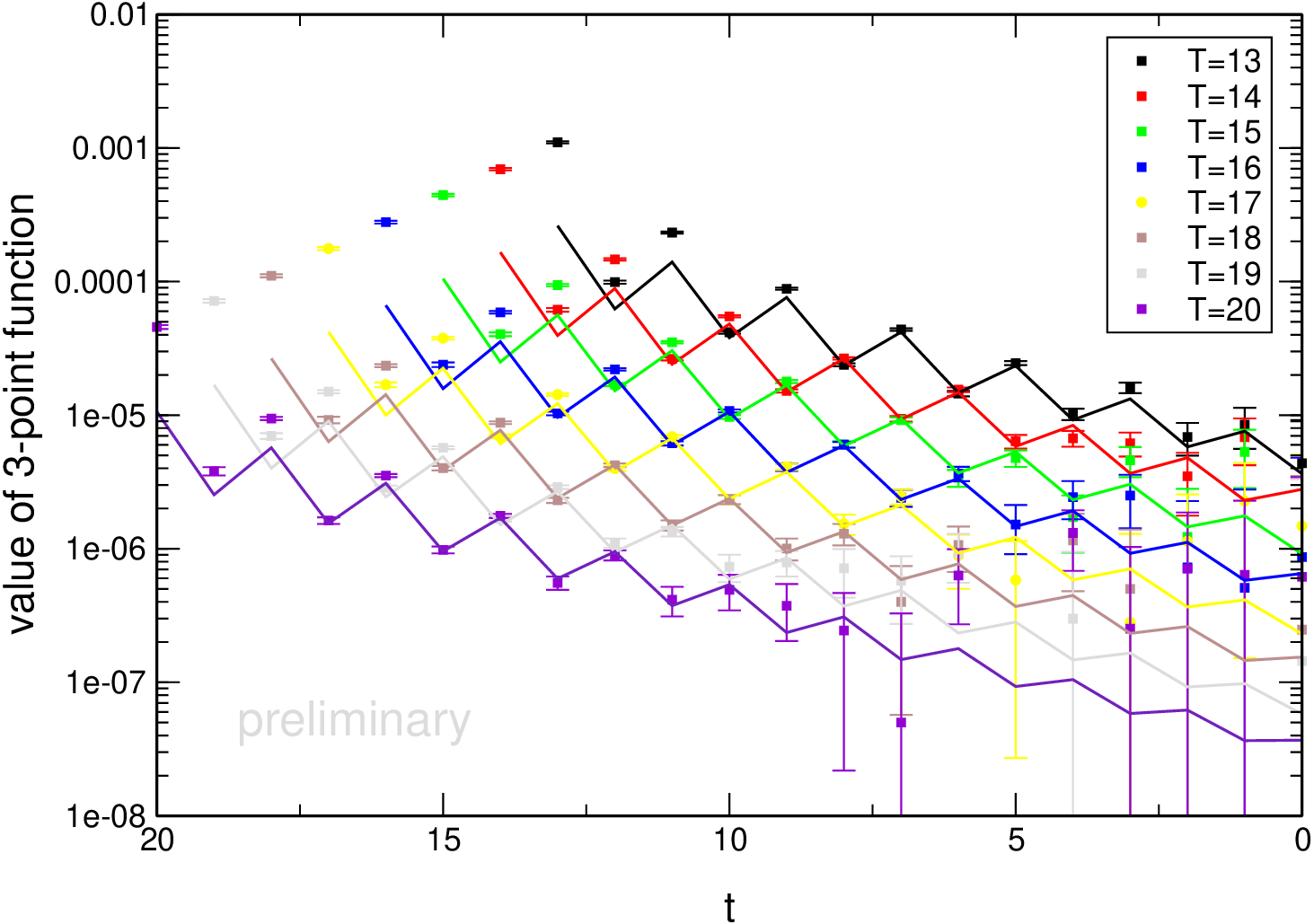}}
\caption{Three-point correlator $\langle \:\:\Phi_K \:\:\: \bar s \gamma_0 b \:\:\:
\Phi_B^\dag\:\: \rangle$ \footnotesize at $\mathbf{k}_{(\mathbf{p})}=\frac{2\pi}{L}(1,0,0)$,
$\mathbf{k}_{(\mathbf{q})}=\frac{2\pi}{L}(2,0,0)$, $\mathbf{v}=(0.4,0,0)$. The fitting range
is $T=14\:...\:18$ and $t=6\:...\:(T-5)$.}
\label{fig:g5_g0_sl_q_2_0_0_p_1_0_0_v_04}
\end{minipage}
\hfill
\begin{minipage}[t]{.48\linewidth}
\centerline{\includegraphics[width=\linewidth]{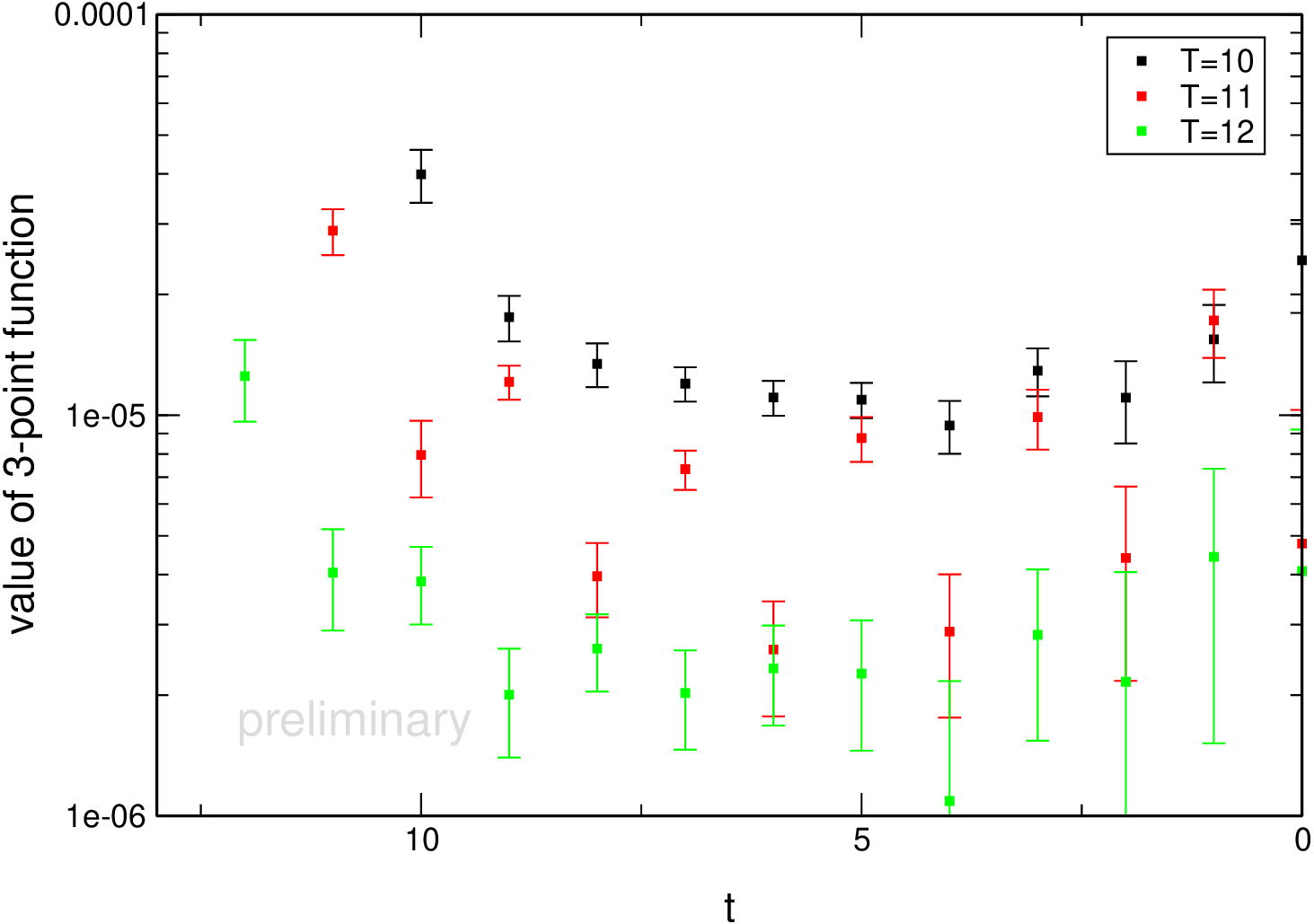}}
\caption{Three-point correlator  $\langle \:\:\Phi_{K^*} \:\:\: \bar s \sigma_{1 3}
b \:\:\: \Phi_B^\dag\:\: \rangle$\footnotesize at
$\mathbf{k}_{(\mathbf{p})}=\frac{2\pi}{L}(1,0,0)$,
$\mathbf{k}_{(\mathbf{q})}=\frac{2\pi}{L}(2,0,0)$, $\mathbf{v}=(0.4,0,0)$.
No reasonable fit was achieved yet.}
\label{fig:g2_s13_sl_q_2_0_0_p_1_0_0_v_04}
\end{minipage}
\end{figure}

To give some examples, plots of the 3-point correlators
$\langle \:\:\Phi_K \:\:\: \bar s \gamma_0 b \:\:\: \Phi_B^\dag\:\: \rangle$ and
$\langle \:\:\Phi_{K^*} \:\:\: \bar s \sigma_{1 3} b \:\:\: \Phi_B^\dag\:\: \rangle$
at the largest $q^2$ (with $\mathbf{v}=0$) are shown in Fig.
\ref{fig:g5_g0_sl_q_0_0_0_p_0_0_0_v_00} and \ref{fig:g2_s13_sl_q_1_0_0_p_0_0_0_v_00}.
The results for the tensor current in combination with the vector meson final state
are much noisier. As expected, the statistical errors are seen to grow further when
the recoil momentum is increased. In Fig. \ref{fig:g5_g0_sl_q_2_0_0_p_1_0_0_v_04}
and \ref{fig:g2_s13_sl_q_2_0_0_p_1_0_0_v_04} we show the corresponding correlators
at $\mathbf{v}=(0.4,0,0)$ and $\mathbf{k}_{(\mathbf{p})}=\frac{2\pi}{L}(1,0,0)$,
$\mathbf{k}_{(\mathbf{q})}=\frac{2\pi}{L}(2,0,0)$. Note that for the fits shown here, 4..6 timeslices from the source/sink were skipped, so that $K=2$, $L=4$ (for the vector final state) or $K=1$, $L=3$ (for the pseudoscalar final state) was sufficient in (\ref{eq:3ptfit}). The results can probably be improved by extending the fitting range and using more exponentials.

\begin{figure}[ht!]
\begin{minipage}[t]{.48\linewidth}
\centerline{\includegraphics[width=\linewidth]{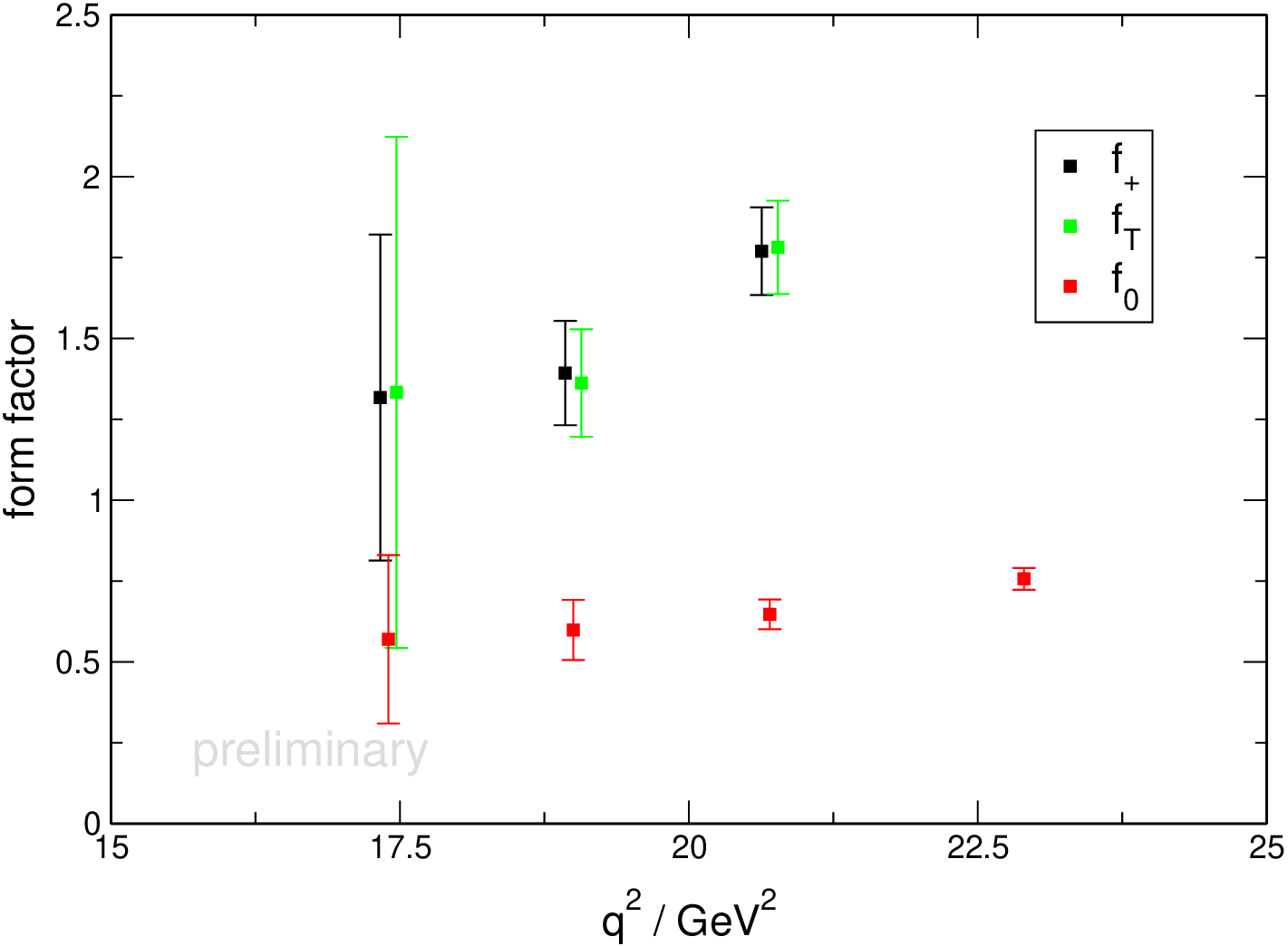}}
\caption{The form factors $f_+$, $f_0$, $f_T$ (points for $f_+$ and $f_T$ are offset
horizontally for legibility). The points at lowest $q^2$ have $\mathbf{v}=(0.4,0,0)$,
$\mathbf{k}_{(\mathbf{p})}=\frac{2\pi}{L}(1,0,0)$,
$\mathbf{k}_{(\mathbf{q})}=\frac{2\pi}{L}(2,0,0)$.}
\label{fig:f0_f+_fT}
\end{minipage}
\hfill
\begin{minipage}[t]{.48\linewidth}
\centerline{\includegraphics[width=\linewidth]{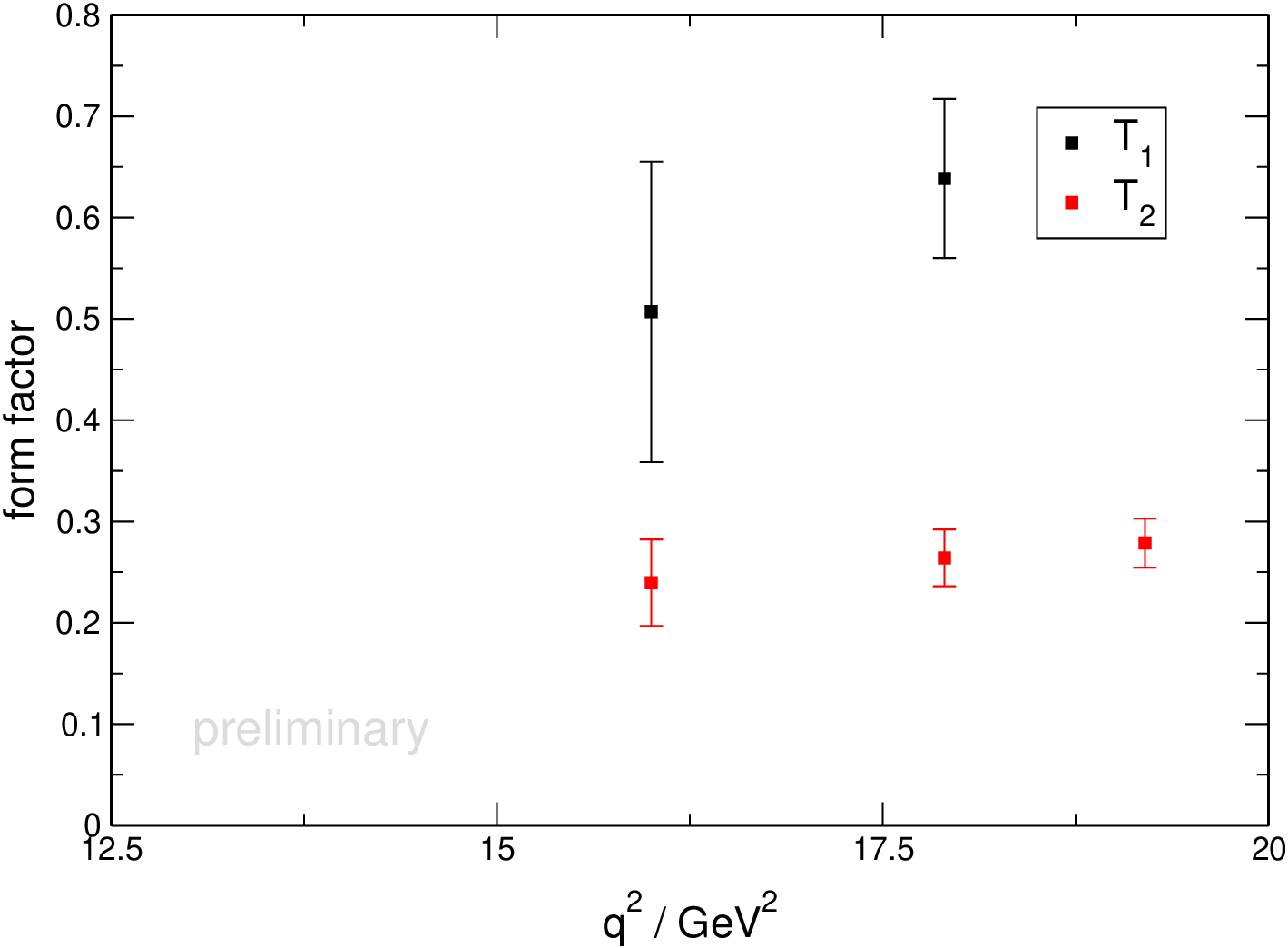}}
\caption{The form factors $T_1$, $T_2$. The points at lowest $q^2$ have
$\mathbf{v}=(0.2,0,0)$, $\mathbf{k}_{(\mathbf{p})}=\frac{2\pi}{L}(1,0,0)$,
$\mathbf{k}_{(\mathbf{q})}=\frac{2\pi}{L}(2,0,0)$.}
\label{fig:T2}
\end{minipage}
\end{figure}

Finally, in Fig. \ref{fig:f0_f+_fT} and \ref{fig:T2} we show some first results for
the form factors $f_0$, $f_+$, $f_T$ and $T_1$, $T_2$. Note that the momentum of the
meson in the final state ($K$ or $K^*$) was exclusively set to the very small values $\mathbf{p'}=0$
or $\mathbf{p'}=\frac{2\pi}{L}(-1,0,0)$. This is made possible by the use of moving NRQCD.


\clearpage

\section{Conclusions}

Using moving-NRQCD and AsqTad actions, we have calculated matching coefficients
for the heavy-light axial-, vector- and tensor currents in the static limit using
1-loop lattice perturbation theory, and performed first computations of form factors
for rare $B$ decays. While moving NRQCD significantly reduces discretisation errors
at low $q^2$, our initial results suffer from large statistical errors, overshadowing
the advantages of the method. However, statistical errors do not constitute a
fundamental obstacle and can be reduced further. The first step will be to extend the
fitting range and include more exponentials. Then, we plan to work with
random-wall sources, which were shown to provide considerable improvement for
semileptonic decays at high recoil momentum \cite{Davies:2007vb}. We will also
use smeared interpolating fields to reduce contributions from excited states, thereby
improving the fits. Furthermore, note that our initial computations were done with
lattice momenta pointing in the 1-direction only. Off-axial momenta and boost velocities
will also allow lower values for $q^2$, for example by using the final meson momentum $\mathbf{p'}=\frac{2\pi}{L}(-1,-1,\:\:0)$.

Once statistical errors are under control, we will study the dependence on the lattice
spacing and the light quark masses. We also plan to include $\mathcal{O}(\Lambda_{QCD}/m_b)$
operators in the matching calculations.

\section*{Acknowledgements}

This work has made use of the resources provided by the Edinburgh Compute and Data Facility
(which is partially supported by the eDIKT initiative), the Fermilab Lattice Gauge Theory
Computational Facility and the Cambridge High Performance Computing Facility.
We thank the MILC collaboration for making their gauge configurations publicly available.

\providecommand{\href}[2]{#2}


\begin{thebibliography}{99}

\bibitem{Dalgic:2006dt}
  E.~Dalgic, A.~Gray, M.~Wingate, C.~T.~H.~Davies, G.~P.~Lepage and J.~Shigemitsu,
  Phys.\ Rev.\  D {\bf 73}, 074502 (2006)
  [Erratum-ibid.\  D {\bf 75}, 119906 (2007)]
  [\href{http://arxiv.org/abs/hep-lat/0601021}{\tt arXiv:hep-lat/0601021}]

\bibitem{Meinel:2007eh}
  S.~Meinel, R.~Horgan, L.~Khomskii, L.~C.~Storoni and M.~Wingate,
  PoS {\bf LAT2007}, 377 (2007)
  [\href{http://arxiv.org/abs/0710.3101}{\tt arXiv:0710.3101 [hep-lat]}].

\bibitem{mNRQCDpaper}
  [HPQCD and UKQCD Collaborations],
  \emph{Moving NRQCD for High Recoil Form Factors in Heavy Quark Physics},
  in preparation.

\bibitem{Wingate:2002fh}
  M.~Wingate, J.~Shigemitsu, C.~T.~H.~Davies, G.~P.~Lepage and H.~D.~Trottier,
  Phys.\ Rev.\  D {\bf 67}, 054505 (2003)
  [\href{http://arxiv.org/abs/hep-lat/0211014}{\tt arXiv:hep-lat/0211014}].

\bibitem{Hart:2004bd}
  A.~Hart, G.~M.~von Hippel, R.~R.~Horgan and L.~C.~Storoni,
  J.\ Comput.\ Phys.\  {\bf 209}, 340 (2005)
  [\href{http://arxiv.org/abs/hep-lat/0411026}{\tt arXiv:hep-lat/0411026}].

\bibitem{Dougall:2004hw}
  A.~Dougall, C.~T.~H.~Davies, K.~M.~Foley and G.~P.~Lepage  [HPQCD
                  Collaboration and UKQCD Collaboration],
  Nucl.\ Phys.\ Proc.\ Suppl.\  {\bf 140}, 431 (2005)
  [\href{http://arxiv.org/abs/hep-lat/0409088}{\tt arXiv:hep-lat/0409088}].

\bibitem{LewThesis}
  L.~Khomskii,
  \emph{Perturbation Theory for Quarks and Currents in Moving NRQCD on a Lattice},
  PhD thesis, in preparation.

\bibitem{Davies:2007vb}
  C.~T.~H.~Davies, E.~Follana, K.~Y.~Wong, G.~P.~Lepage and J.~Shigemitsu,
  PoS {\bf LAT2007}, 378 (2007)
  [\href{http://arxiv.org/abs/0710.0741}{\tt arXiv:0710.0741 [hep-lat]}].

\end{thebibliography}
\end{document}